\documentclass{PoS}
\usepackage[font=footnotesize,labelfont=bf,labelsep=period]{caption}
\usepackage[numbers,sort&compress]{natbib}
\usepackage{setspace}
\usepackage{amsmath,amssymb,amsbsy,booktabs}

\newcommand{\txt}[1]{\mathsf{#1}}
\providecommand{\abs}[1]{\lvert#1\rvert}    

\title{Neutral $B$-meson mixing from dynamical lattice QCD with chiral light quarks and static $b$-quarks}

\ShortTitle{Neutral $B$-meson mixing from dynamical lattice QCD}

\author{\speaker{Oliver Witzel} \vspace{4mm}\newline  for the RBC and UKQCD collaborations\vspace{4mm}\\
        Department of Physics, Brookhaven National Laboratory, Upton, NY 11973, USA\\
        E-mail: \email{witzel@quark.phy.bnl.gov}}

\abstract{In the limit of infinitely heavy $b$-quarks we compute the $SU(3)$-breaking ratio $\xi$ of neutral $B$-meson mixing matrix elements. We also present results for the ratio of decay constants $f_{B_s}/f_{B_d}$. Our calculation employs chirality-preserving domain-wall fermions for the light quarks, a static 
action with link-smearing for the $b$-quarks, and the Iwasaki gauge action. 
Here we report on our results from the $16^3 \times 32 \times 16$ ensemble ($a^{-1} = 1.729(28)$ GeV) which we 
use to verify our method.  We improve upon our earlier work by including 
$O(\alpha_s pa)$ matching for the computation of the decay constants and 
extrapolating to the physical point using chiral perturbation theory.}

\FullConference{The XXVII International Symposium on Lattice Field Theory - LAT2009\\
		 July 26-31 2009\\
		 Peking University, Beijing, China}

\begin{document}

\section{Introduction} 
\begin{figure}[b]
\begin{center}
\vspace{-3mm}
\includegraphics[width=0.8\textwidth]{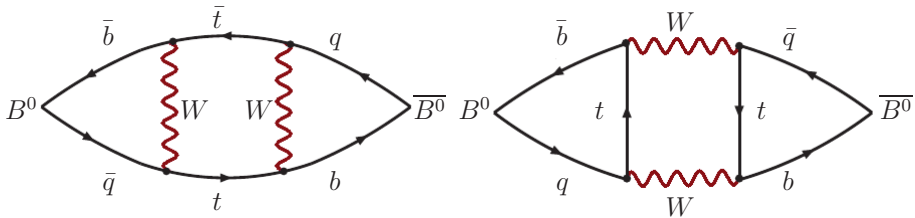}
\caption{The dominant contribution to $B$-$\bar B$-mixing.}
\end{center}
\label{fig:MixingDiagrams}
\end{figure}

\vspace{-2mm}
Neutral $B$-$\bar B$-mixing is a phenomenologically important quantity because it allows us to determine CKM matrix elements.  Within the Standard Model the dominant contribution is given by box diagrams with top quarks as shown in Figure \ref{fig:MixingDiagrams}.  Experimentally,  the observable quantity is the mass difference (also named oscillation frequency) $\Delta m_q$, where the subscript $q$ labels the light quark content ($d$ or $s$) of the $B$ meson.  The mass difference is parametrized as \cite{Buras:1990fn}
\begin{align} 
\Delta m_q = \frac{G_F^2m^2_W}{6\pi^2} \eta_B S_0 m_{B_q}{f_{B_q}^2B_{B_q}} \abs{V_{tq}^*V_{tb}}^2.
\end{align}
where $m_{B_q}$ is the mass of the $B_q$-meson, and  $V_{tq}^*$ and $V_{tb}$ denote CKM matrix-elements. The Inami-Lim function, $S_0$ \cite{Inami:1980fz}, and the QCD coefficient, $\eta_B$ \cite{Buras:1990fn}, can be computed perturbatively, while $f_q^2 B_{B_q}$ is the non-perturbative input: the decay constant $f_q$ and the $B$-meson bag parameter $B_{B_q}$.

We define $SU(3)$ breaking ratios as the ratio of a quantity for the $B_s$-meson over the same quantity for the $B_d$ -meson. In particular we are interested in 
\begin{align}
\xi &= \frac{f_{B_s}\sqrt{B_{B_s}}}{f_{B_d}\sqrt{B_{B_d}}}
\end{align}
and we also consider the ratio of the decay constants ${f_{B_s}}/{f_{B_d}}$.  Computing $\xi$ non-perturbatively allows one with additional experimental input to extract the ratio of CKM matrix elements 
\begin{align}
\frac{\Delta m_s}{\Delta m_d} = \frac{m_{B_s}}{m_{B_d}}\,{\xi^2} \, \frac{\abs{V_{ts}}^2}{\abs{V_{td}}^2}.
\end{align}

The phenomenological importance of $B$-$\bar B$ mixing is given by the fact that the ratio ${\abs{V_{ts}}^2}/{\abs{V_{td}}^2}$ constrains the apex of the unitarity triangle \cite{Antonelli:2009ws}.  Experimentally, $\Delta m_d$ and $\Delta m_s$ are measured to better than a percent \cite{Amsler:2008zzb,Abazov:2006dm,Abulencia:2006mq,Abulencia:2006ze}, whereas we know $\xi$ only to about $3\%$. Hence in order to get a stronger constraint on the apex of the unitarity triangle and consequently on new physics beyond the Standard Model, we need to improve the determination of $\xi$.

The project presented here aims at this task by demonstrating the viability of our method on $16^3\times 32 \times 16$ dynamical domain-wall fermion ensembles generated by the RBC and UKQCD collaborations \cite{Allton:2007hx}.  Our central values agree with the ones published in the literature, however our errors are large. A future follow-up project will hopefully lead to results with comparable uncertainties to those obtained by Gamiz et al.~\cite{Gamiz:2009ku} and Evans et al.~\cite{ToddEvans:2008}. Details of the present computation will soon be published in \cite{ThePaper}.

\section{Actions and parameters} \label{sec:Method}\vspace{-2mm}
In order to accomplish the computation of $B$-$\bar B$-mixing we choose domain-wall fermions \cite{Kaplan:1992bt,Shamir:1993zy} for the light quarks ($u$, $d$, $s$), the Iwasaki gauge action \cite{Iwasaki:1983ck}, and a static action with link-smearing for the $b$-quarks \cite{Eichten:1989kb}.  

Domain-wall fermions are formulated in five dimensions and posses an approximate chiral symmetry. Left-handed modes are bound to a 4-d brane at $s=0$, while right-handed modes are bound to a 4-d brane at $s=L_s-1$. The overlap between both is exponentionally suppressed. Besides the approximate chiral symmetry, domain-wall fermions are advantageous because the renormalization is simplified due to reduced operator mixing.  The choice of the Iwasaki gauge action is motivated by the fact that it improves chiral symmetry and reduces the residual quark mass in combination with domain-wall sea quarks \cite{Aoki:2002vt} over e.g.~the Wilson gauge action \cite{Wilson:1974sk}.

Our simulations use  $16^3 \times 32 \times 16$ dynamical 2+1 flavor domain-wall lattices \cite{Allton:2007hx} in which the up and down sea quarks are degenerate and the strange sea quark is slightly heavier than its physical value \cite{Allton:2008pn}. The domain-wall height is set to $M_5 = 1.8$ and we have $\beta = 2.13$. We estimate the inverse lattice spacing to be $a^{-1} = 1.729(28)$ GeV and the residual quark mass to be $a m_\text{res}=0.00315$ \cite{Allton:2008pn}.  Hence $a$ is approximately $0.11$ fm and we have a (1.8 fm)$^3$ box.  In Table \ref{tab:SimParameters} we list the values of the light ($m_l$) and heavy ($m_h$) sea quark masses, the values of the valence quark masses ($m_x$), the mass of the pions, and the number of configurations used in our simulations.

\begin{table}[bt]
\begin{center}
\begin{tabular}{cccccc}\toprule
        &         &         &             &\multicolumn{2}{c}{\# configs.}\\
$a m_l$ & $a m_h$ & $a m_x$ & $m_\pi$ (MeV)& APE & HYP \\ \midrule
0.01    & 0.04    & 0.01, 0.0359 & 400 & 298& 300\\
0.02    & 0.04    & 0.02, 0.0359 & 530 & 298& 300\\
0.03    & 0.04    & 0.03, 0.0359 & 630 & 298& 300\\ \bottomrule
\end{tabular}
\caption{The light ($m_l$) and heavy ($m_h$) sea quark masses, the valence quark masses ($m_x$), the mass of the pions ($m_\pi$) and the number of configurations used in our simulations.}
\label{tab:SimParameters}
\end{center}\vspace{-5mm}
\end{table}

For the $b$-quarks we use a static action which is based on the original lattice formulation by Eichten and Hill \cite{Eichten:1989kb} but improved by link-smearing.  This formulation corresponds to an effective action in the limit of infinitely heavy $b$-quarks, i.e.~corrections of order $1/m_b$ are neglected. Advantages of this static action are that the static quark propagator is given as product of gauge links, the propagator is $O(a)$ improved  and has a simple continuum limit. We also enhance the signal-to-noise ratio by link-smearing in comparison to the original formulation \cite{DellaMorte:2003mn}. For this work we utilize APE smearing \cite{Albanese:1987ds, Falcioni:1984ei} and hypercubic blocking (HYP smearing) \cite{Hasenfratz:2001hp, DellaMorte:2005yc}.

In case of APE smearing with smearing parameter $\alpha=1$, one computes all staples for a given link, adds them and projects the result finally back onto $SU(3)$.  HYP smearing consists of three steps of APE smearing restricted to the links within the hypercube surrounding a given link. For each step $i$ we have a choice for the smearing parameters $\alpha_i$. Following Della Morte et al.~\cite{DellaMorte:2005yc} we choose $(\alpha_1, \alpha_2, \alpha_3)=(1.0,\, 1.0,\, 0.5)$, which is commonly referred to as ``HYP2''.  The required $SU(3)$ projection is not unique: in the case of APE smearing it is performed by the unit circle projection method \cite{Kamleh:2004xk}, while for HYP smearing we employ an iterative procedure which yields the $SU(3)$ projection of a matrix $V$ by seeking $U_\txt{max} \in SU(3)$ such that $\txt{Re}\, \txt{Tr}(U_\txt{max}V^\dagger)$ is maximal \cite{Bali:1992ab}. Both projections are equivalent in the weak coupling limit \cite{Kamleh:2004xk, ThePaper}.

\section{Lattice calculation} \label{sec:LatticeRenormalizationChiral}\vspace{-2mm}
The perturbative renormalization of the heavy-light axial current and $\Delta B = 2$ four fermion operator follow a two-step matching procedure. First we match the continuum QCD operators at a scale $\mu_b$ in the $\overline{MS}$-scheme using naive dimensional regularization onto operators in the continuum static effective theory at a scale $\mu$. Next, we match the operators in the continuum static effective theory to operators on the lattice \cite{Talk:Taku}. Combining the results of the two matching steps we obtain the perturbatively computed matching coefficients $c_A$, $Z_\Phi$, $Z_{VA}$, and $Z_{SP}$, which are listed in Table \ref{tab:Matching} \cite{PT_Oa}. We account for the truncation of the perturbative series in our estimate of the systematic errors. 

\begin{table}[bt]
\begin{center}
\begin{tabular}{ccccc}
\toprule
 smearing &  $c_A$ &$Z_\Phi$  &  $Z_{VA}$ & $Z_{SP}$ \\ \midrule
APE & 0.0653 & 0.9507 & 0.7485 & -0.1448 \\
HYP & 0.1204 & 0.9813 & 0.8108 & -0.1448 \\
\bottomrule
\end{tabular}
\caption{Perturbative matching coefficients for bilinear and four-quark operators evaluated for APE- and HYP-smeared static-quark gauge links.}
\label{tab:Matching}
\end{center}\vspace{-5mm}
\end{table}

For the bilinear operators we include the $O(\alpha_s pa)$ improvement term in our analysis
\begin{align}
\Phi^\text{ren}_B = Z_\Phi(1+c_A \sinh(m^*_B))\Phi_B^\text{lat},
\end{align}
but we have not yet implemented $O(pa)$ improvement for the four-quark operator. The renormalized decay amplitude $\Phi^\text{ren}_B$ is related to the decay constant $f_{B_q} = \Phi^\text{ren}_B/\sqrt{m_{B_q}}$ and is obtained by computing on the lattice the ratio of two-point functions of the heavy-light axial current using local (L) and wall (W) sources/sinks:
\begin{align}
\Phi_B^\text{lat} = \lim_{t\gg t_0}\sqrt{\frac{2}{L^3}}\frac{\abs{{\cal C}^{LW}(t,t_0)}}{\sqrt{{\cal C}^{WW}(t,t_0)\text{e}^{-m_{B_q}^*(t-t_0)}}},
\end{align}
where $m_{B_q}^*$ is the unphysical $B_q$-meson rest mass. For the matrix elements containing the four-fermion operator we compute additional three-point functions ${\cal C}_{\cal O}$ using box sources/sinks (B) \cite{Christ:2007cn} in case of APE smearing:
\begin{align}
M_{\cal O}^\text{lat} = \lim_{t_f\gg t \gg t_0}2\, \frac{{\cal C}_{\cal O}^B(t_f,t, t_0) \text{e}^{m_{B_q}^*(t_f-t_0)/2}}{\sqrt{{\cal C}^{BB}(t,t_f){\cal C}^{BB}(t,t_0)}},
\end{align}
 and  wall sources/sinks (W) for HYP smearing:
\begin{align}
M_{\cal O}^\text{lat} = \lim_{t_f\gg t \gg t_0} L^3 \frac{{\cal C}_{\cal O}^W(t_f,t, t_0)}{{\cal C}^{LW}(t,t_f){\cal C}^{LW}(t,t_0)} \cdot\left( \Phi_{B_q}^\text{ren}\right)^2,
\end{align}
Because the four-quark operators of different chiralities mix under renormalization, we obtain the renormalized expression for the matrix element by
\begin{align}
M^\text{ren}_{B_q} = Z_{VA} M^\text{lat}_{VV+AA} + Z_{SP}M^\text{lat}_{SS+PP}.
\end{align}

Finally, we extrapolate our lattice data to the physical quark masses and the continuum using next-to-leading order partially quenched $SU(3)$ heavy-light meson chiral perturbation theory.  Schematically, the expressions for the $SU(3)$ breaking ratios are given by
\begin{align}
\frac{\Phi_{B_{s^\prime}}^\textrm{ren}}{\Phi_{B_l}^\textrm{ren}}  & =  1 + \textrm{``chiral logs"} + \frac{2\mu}{(4\pi f)^2}\tilde{c}_\textrm{val} (m_{s^\prime} - m_l), \\
\sqrt{\frac{M^\textrm{ren}_{B_{s^\prime}}}{M^\textrm{ren}_{B_l}}}  & =  1 + \textrm{``chiral logs"} + \frac{\mu}{(4\pi f)^2} \tilde{d}_\textrm{val} (m_{s^\prime} - m_l),
\end{align}
where the quark masses are expressed as dimensionless ratios and ``chiral logs'' denote non-analytic functions of the pseudo-Goldstone meson masses. Performing a linear, one-parameter fit of our data with respect to the expressions above we are able to extract the physical value for $\Phi^\txt{ren}_s/\Phi^\text{ren}_d$ and $\xi$. As input parameters we use: $\Lambda_\chi = 1$ GeV, $\mu = 2.35(16)$ \cite{Allton:2008pn}, $f_\pi = 130.4$ MeV \cite{Amsler:2008zzb},  $g_{B^*B\pi} = 0.516$ \cite{Ohki:2008py},  $a m_{ud}+ am_{res} = 0.001300(62)$, $a m_s+a m_{res}   = 0.0375(17)$ \cite{Allton:2008pn}.  A detailed description of our chiral extrapolation will be included in \cite{ThePaper}, where we also will discuss the alternative of using heavy-light $SU(2)$ chiral perturbation theory.

\section{Results  and Conclusion} \label{sec:Results}
\begin{figure}[tb]
\begin{picture}(150,54)
\put(0,0){\includegraphics[width=0.52\textwidth,clip]{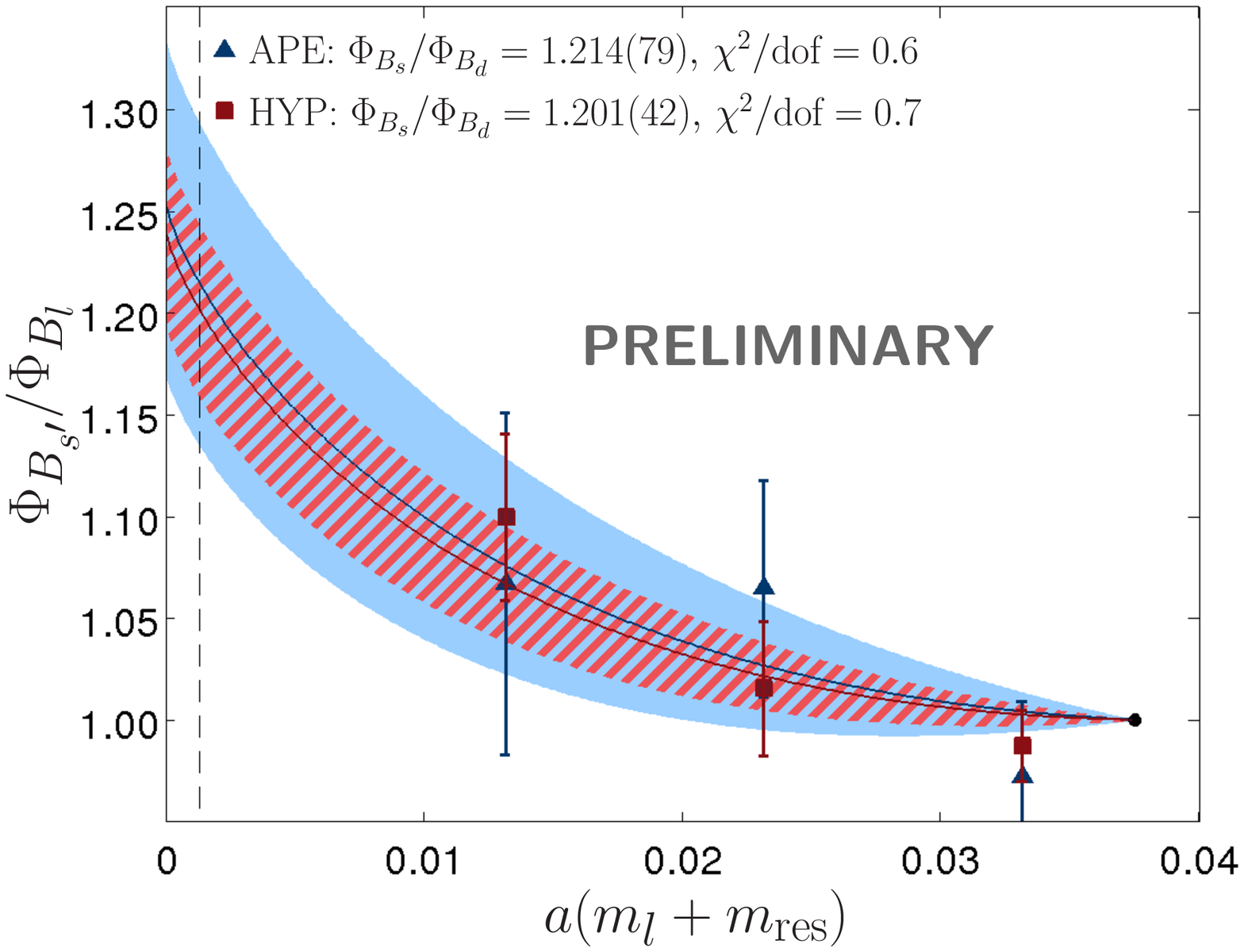}}
\put(75,0){\includegraphics[width=0.52\textwidth,clip]{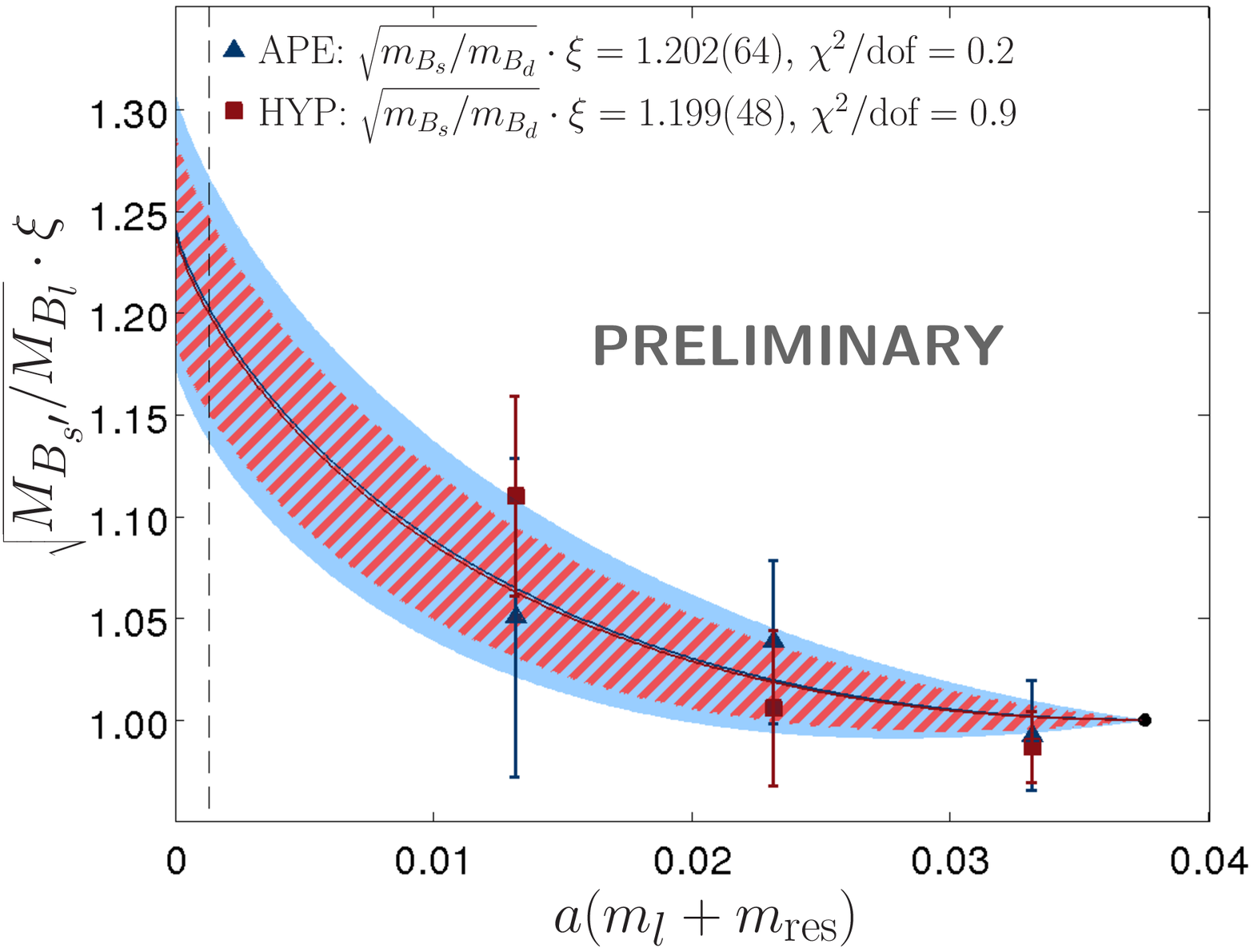}}
\end{picture}
\caption{Chiral extrapolation of $\Phi_{B_s}/\Phi_{B_d}=\sqrt{m_{B_s}/m_{B_d}}\cdot f_{B_s}/f_{B_d}$ (left) and $\sqrt{m_{B_s}/m_{B_d}} \cdot \xi$ (right). The APE data points are shown as blue triangles, whereas the HYP data points are shown as red squares. The color of the shaded (hatched) error bands match those of the APE (HYP) data points. The physical average $u-d$ quark mass is indicated by the dashed line and the black dot denotes the physical strange quark mass.  Only statistical errors are shown.}
\label{fig:Fits}
\end{figure}

\vspace{-2mm}
We present the preliminary results of our chiral extrapolation in Fig.~\ref{fig:Fits}. The plot on the left shows the data and the fit for $\Phi_{B_s}/\Phi_{B_d}$, the plot on the right the outcome for $\sqrt{m_{B_s}/m_{B_d}}\cdot \xi$. In both cases only statistical errors are shown, which are computed following to Ref.~\cite{Wolff:2003sm}. In addition we estimate systematic errors by varying the input parameters around their uncertainty, considering a constrained linear fit as alternative fit function, and using power-counting for the discretization errors, the errors due to the renormalization factors, the finite volume errors $1/m_b$ corrections. All of these errors are listed in Table \ref{tab:TotalError} and discussed in detail in \cite{ThePaper}.

\begin{table} 
\begin{center}\vspace{-5mm}
\begin{tabular}{lcccccc} \\ \toprule

& &\multicolumn{2}{c}{$f_{B_s}/f_{B_d}$} & &\multicolumn{2}{c}{$\xi$} \\
uncertainty                  	 		& \quad\quad &APE & HYP & \quad\quad & APE & HYP \\ \midrule
statistics           				&& 7\% & 4\% && 5\%& 4\% \\ \hline
chiral extrapolation       			&& 11\% & 11\% && 12\%& 11\% \\
uncertainty in $g_{B^*B\pi}$ 		&& 4\% & 4\%&& 3\%&  3\% \\
discretization error       			&& 3\% & 3\%&& 4\% & 4\%\\
renormalization factors    			&& 0\% & 0\%&& 2\% & 2\% \\
scale and quark mass uncertainties && 1\%& 1\% && 1\% & 1\% \\
finite volume error       			&& 1\% & 1\%&& 1\% & 1\%\\
$1/m_b$ corrections        			&& 2\% & 2\%&& 2\% & 2\%\\
\hline
total systematics       			&& 12\% & 12\%&& 13\%& 12\%\\
\bottomrule
\end{tabular}
\caption{Total error budget for the $SU(3)$-breaking ratios $f_{B_s}/f_{B_d}$ and $\xi$ rounded to the nearest percentage.}
\label{tab:TotalError}
\end{center}\vspace{-6mm}
\end{table}

\begin{figure}[b]
\vspace{-4mm}
\includegraphics[width=\textwidth]{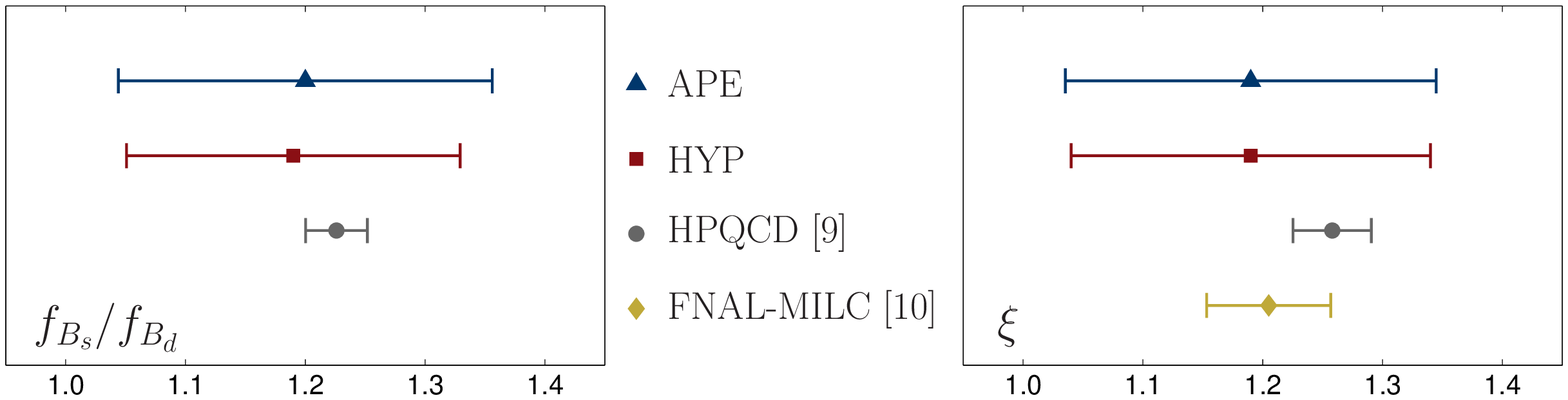}
\caption{Comparison of lattice QCD results for $f_{B_s}/f_{B_d}$ (left) and $\xi$ (right). Our new preliminary data points are marked as blue triangle (APE) and red square (HYP). The gray circle marks the values published by the HPQCD collaboration \cite{Gamiz:2009ku}, the beige diamond the preliminary value presented by the FNAL-MILC collaboration at Lattice 2008\cite{ToddEvans:2008}.}
\label{fig:LatResults}
\end{figure}

Finally, we use the experimentally-measured ratio of the masses $m_{B_s^0}/m_{B_d^0} = 5366.6 / 5279.5 = 1.0165$~\cite{Amsler:2008zzb} to obtain the following values for the $SU(3)$-breaking ratios of $B$-meson decay constants and mixing matrix elements:
\begin{align}
\frac{f_{B_s}}{f_{B_d}} = 
\left\{ \begin{aligned}
& 1.20(08)(14) \quad\textrm{APE}\\
& 1.19(05)(14) \quad\textrm{HYP}
\end{aligned}\right. \, ,
%
\qquad \text{and} \qquad
\xi =
\left\{ \begin{aligned}
&1.19 (06)(15) \quad\textrm{APE}\\
&1.19 (05)(14) \quad\textrm{HYP}
\end{aligned}\right. \, ,
\end{align}
where the first errors are statistical, the second are the sum of all systematic errors added in quadrature. Currently, we are updating our analysis and hence these values may change in our publication \cite{ThePaper}.  We find that both smearings used agree very well indicating that the discretization errors are small in the ratios. When comparing these new results to the ones in the literature published by the HPQCD collaboration \cite{Gamiz:2009ku} and presented by the FNAL-MILC collaboration at Lattice 2008 \cite{ToddEvans:2008}, we also find good agreement (see Fig.~\ref{fig:LatResults}). However, our errors are large and we look forward to improve upon them in future works.  \vspace{-2mm}

\section*{Acknowledgments}\vspace{-2mm}
I am thankful to all the members of the RBC and UKQCD 
collaborations.  
Numerical computations for this work were performed on 
the QCDOC computers of the RIKEN-BNL Research Center 
and the USQCD Collaboration, in part funded by the 
Office of Science of the U.S.~Department of Energy.
This manuscript has been authored by 
an employee of Brookhaven Science Associates, LLC 
under Contract No.~DE-AC02-98CH10886 with the 
U.S.~Department of Energy.

\begin{spacing}{0.5}
\bibliography{B_meson}
\bibliographystyle{apsrev4-1}
\end{spacing}
\end{document}